\documentclass[journal=ascefj,manuscript=article,layout=singlecolumn]{achemso}

\usepackage[version=3]{mhchem} 

\usepackage{bm,mathtools,amsmath,amssymb,amsbsy}
\usepackage{latexsym,mathrsfs,enumerate,color}
\usepackage[mathcal]{euscript}
\usepackage{upgreek}
\usepackage{graphicx}
\usepackage{xifthen}
\usepackage[most]{tcolorbox}
\usepackage{pdfpages}

\author{Wenya Song}
\affiliation{Helmholtz-Zentrum Dresden-Rossendorf e.V. (HZDR), Institute of Ion Beam Physics and Materials Research, Bautzner Landstr. 400, 01328 Dresden, Germany}
\author{Gungun Lin}
\affiliation{Helmholtz-Zentrum Dresden-Rossendorf e.V. (HZDR), Institute of Ion Beam Physics and Materials Research, Bautzner Landstr. 400, 01328 Dresden, Germany}
\alsoaffiliation{Institute for Biomedical Materials and Devices (IBMD) School of Mathematical and Physical Sciences, Faculty of Science, University of Technology Sydney, Ultimo NSW 2007 Australia}
\email{gungun.lin@uts.edu.au}
\author{Jin Ge}
\affiliation{Helmholtz-Zentrum Dresden-Rossendorf e.V. (HZDR), Institute of Ion Beam Physics and Materials Research, Bautzner Landstr. 400, 01328 Dresden, Germany}
\author{J{\" u}rgen Fassbender}
\affiliation{Helmholtz-Zentrum Dresden-Rossendorf e.V. (HZDR), Institute of Ion Beam Physics and Materials Research, Bautzner Landstr. 400, 01328 Dresden, Germany}
\alsoaffiliation{Technische Universit{\" a}t Dresden, Zellescher Weg 16, 01069 Dresden, Germany}
\author{Denys Makarov}
\affiliation{Helmholtz-Zentrum Dresden-Rossendorf e.V. (HZDR), Institute of Ion Beam Physics and Materials Research, Bautzner Landstr. 400, 01328 Dresden, Germany}
\email{d.makarov@hzdr.de}

\title {Encoding micro-reactors with droplet chains in microfluidics}


\begin{document}


\begin{abstract}
  Droplet-based high throughput biomolecular screening and combinatorial synthesis entail a viable indexing strategy to be developed for the identification of each micro-reactor. Here, we propose a novel indexing scheme based on the generation of droplet sequences on demand to form unique encoding droplet chains in fluidic networks. These codes are represented by multiunit and multilevel droplets packages, with each code unit possessing several distinct signal levels, potentially allowing large encoding capacity. For proof of concept, we use magnetic nanoparticles as the encoding material and a giant magnetoresistance (GMR) sensor-based active sorting system supplemented with an optical detector to generate and decode the sequence of one exemplar sample droplet reactor and a 4-unit quaternary magnetic code. The indexing capacity offered by 4-unit multilevel codes with this indexing strategy is estimated to exceed 104, which holds great promise for large-scale droplet-based screening and synthesis. 

\end{abstract}


Femto-to picoliter monodisperse emulsion droplets can be generated by injecting an aqueous fluid phase into an immiscible phase in channels with submillimeter dimensions, or vice versa \cite{Garstecki2006, Schneider2011}. The enormous interest in fully exploiting the potential of emulsion droplets as micro-reactors with controlled volumes has led to the emergence of a vibrant research field, namely, ‘droplet microfluidics’ \cite{Teh2008, Clausell2010, Koehler2013, Xi2017, Kaminski2017}. In recent years, droplet-based microfluidic devices have been extensively studied for generic combinatorial synthesis of biological and chemical materials \cite{Knauer2012, Bawazer2016}, drug screening \citep{Miller2012}, and high-throughput multiplexed biological assays \cite{Baraban2011, Guo2012, Zhu2016, Karnaushenko2015, Chang2015, Lange2016}.

\begin{figure}
	\includegraphics[width=\linewidth]{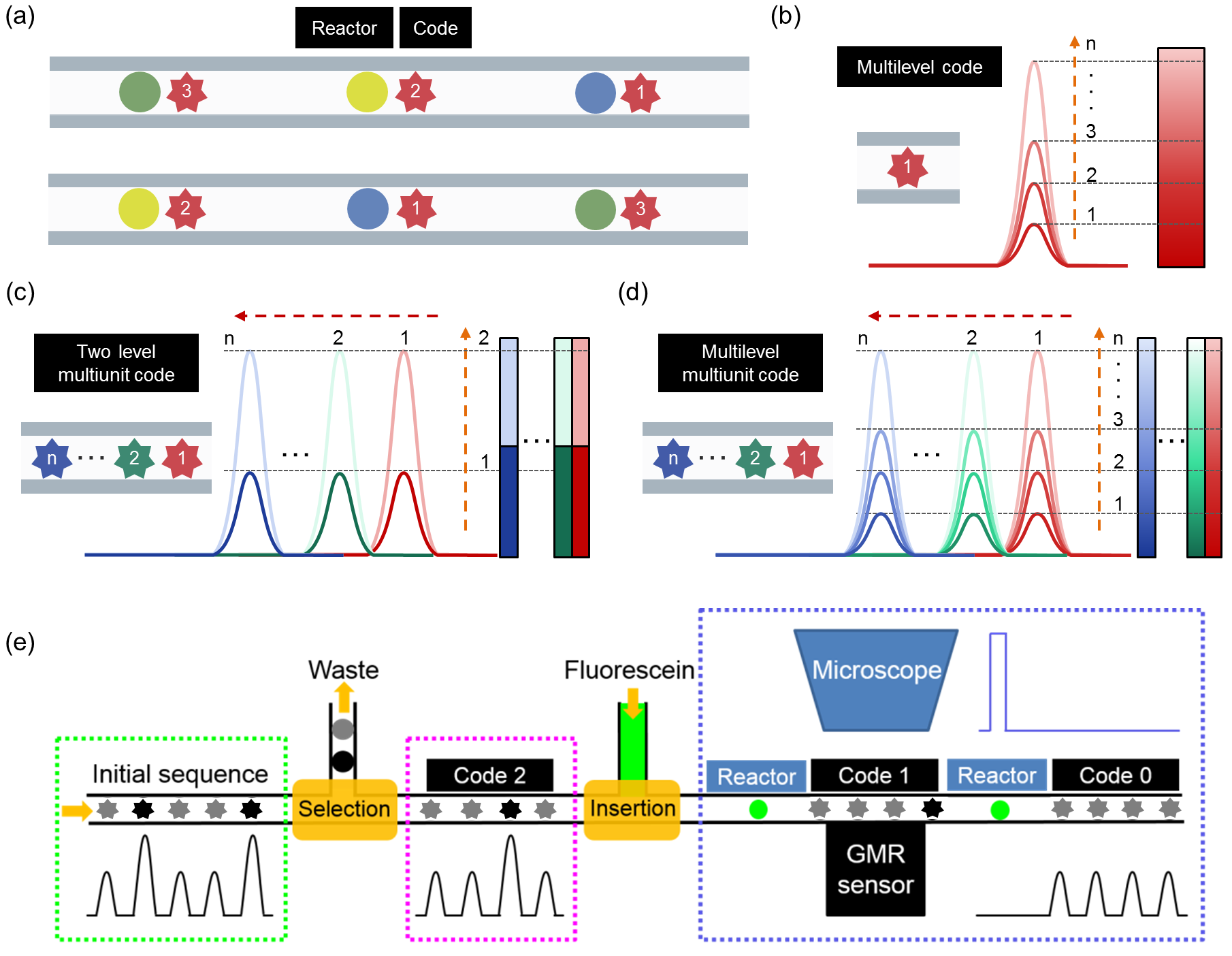}
	\caption{(a) Encoding droplet reactors with a neighboring droplet code. A code is denoted by a star shape and a droplet reactor is denoted by a filled circle. Schematics of (b) a single unit code with multiple levels, (c) multiunit codes with two levels, and (d) multiunit code with multiple levels. (e) Schematic of the operation principle of code creation and verification in fluidics. The selection and detection of the signals of an initial sequence and the verification of created codes are conducted with GMR sensors. The entirety of a magnetic code and a reaction droplet is detected by a GMR sensor sensing system and a high-speed camera fitted to a far-field microscope, respectively. Black and gray star symbols indicate droplets with high and low concentrations of MNPs. Green circles indicate fluorescein droplets.}
    \label{fig:figure1}
\end{figure}

A grand vision of developing droplet microfluidic platforms is that once a large number of droplets could encapsulate various samples, they being operated with ease and identified with high fidelity, massively parallelized reactions or screening could be anticipated. For this vision, an encoding strategy is essential to index droplets carrying a variety of samples \cite{Pompano2011, Knauer2012}. So far, the most commonly employed encoding approach have been the co-loading of samples and encoding materials within the same droplets \cite{Miller2012}, having the possibility of realizing multistep synthesis in a fully automatic way (a possible use case is sketched in Figure S1). A variety of encoding materials such as fluorescent dyes, quantum dots and magnetic particles is available. These encoding materials enable the generation of unique droplet identifiers that are embodied by discrete fluorescent intensity levels, wavelengths or magnetic signal patterns \cite{Ding2013, Lin2015}. The unique encoded information can be decoded by optical and magnetic detectors, respectively. A general strategy of increasing encoding capacity is by increasing the dimensionality of codes, such as combining fluorescent intensity detected by means of optics with other e.g. all electric sensors, offering alternative encoding methods \cite{Chen2004, Lin2015, Karnaushenko2015, Schuett2016}. One of the attractive possibilities is given by a conjunction of the fluorescent and magnetic encoding capacities \cite{Lin2015}. However, with more colors and bulky optical components such as excitation sources and filters are required. The combination of optical and magnetic codes, though bearing the advantages of boosting the encoding capacity, is still at the expense of the cost and complexity of the whole system \cite{Lin2015}. Furthermore, a key drawback of current droplet encoding strategy is that the encoding materials are typically loaded with samples in the same droplets, which may cause undesirable interference to the analyses of cell behaviors for drug screening due to the different concentrations of coencapsulated encoding materials. There is also a possibility of nanocrystals to be synthesized being contaminated by the unwanted encoding materials. Under such circumstances, it is crucial to separating samples and encoding materials. A solution is by using a droplet neighboring to a sample-containing droplet as a code, as shown in Figure~\ref{fig:figure1}a. This approach is promising to overcome the drawbacks of conventional encoding strategy without the concern of altering the initial composition of the samples. So far, such a position-encoding strategy was demonstrated by \textit{passively} producing alternating sample-containing and fluorescent dye-containing droplets with a pair of T-junctions fabricated in microfluidic channels \cite{Zheng2004}. However, this method offers little flexibility to adjust the number of droplets near the sample droplets. Hence, the current encoding capacity of single code with multiple levels (Figure~\ref{fig:figure1}b) and multiunit code with two levels (Figure~\ref{fig:figure1}c) is still limited for encoding in a single droplet. 

Here, we put forth the concept of \textit{multiunit multilevel codes} (Figure~\ref{fig:figure1}d) potentially allowing to achieve a large indexing capacity without worrying the alteration of initial compositions of samples. In this concept, unique sequences of droplets are created on demand in microfluidic networks as codes, of which each elementary droplet unit exhibits a distinct signal level. We demonstrate 4-unit quaternary codes which could potentially offer a capacity of 256, which are 64 times larger than that of single-unit codes with the  same number of intensity levels applied by scaling up the encoding capacity without scaling up the device. 

For proof-of-concept demonstration, we use ferrofluid magnetic nanoparticles (MNPs) as encoding materials to be encapsulated in droplets (Figure~\ref{fig:figure1}e). A modular fluidic code generation system is established and armed with a magnetic field sensor and electro-magnetic valves. We operate the device to create a sequence of reaction droplets encoded by 4-unit quaternary codes. The droplet reactors are then decoded using the magnetic detector. To demonstrate the potential of the encoding strategy for multiplexed assays and synthesis, we exemplarily load fluorescent dyes (green color) in droplets between the encoding droplet sequences to mimic the products to be analyzed/synthesized in reaction droplets. In this way, the reaction products can be monitored by optically observing the fluorescence. By combining the intensity levels, the limit of the indexing capacity offered by the 4-unit codes using the setup is explored to reach 38416, which is sufficient to perform the indexing task for producing large size compound libraries for drug screening.

\section{Results and Discussions}

\subsection{Concept of droplet-based  positional encoding}

Our concept of multiunit multilevel positional encoding for droplet fluidics is based on an active droplet selection strategy, which can be realized by a fluidic system with an encoded signal detector, valves and a rigorous droplet selection algorithm. A code is embodied by a chain of droplets consisting of multiple units exhibiting differentiable signal levels. Each droplet unit is physically isolated out of an initial droplet chain produced from the upstream of a fluidic network, while the number of the units forming the code and the encoded signal levels of individual units are pre-defined by a computer program. This functionality is relying upon the active switch of a 3-way valve, one way connected to the up-stream channel, the others connected to down-stream channels diverting selected individual droplet units for sample indexing and remaining droplet units for recycling. The mechanical activation of the valve is controlled by a feedback from the signal detector which analyses the signal of each interrogated droplet. Right after generating a desired code, a droplet reactor to be indexed is inserted behind the code for indexing. In practice, such a concept could be applicable in a variety of detectors and encoding signals combinations including optical detectors and fluorescent signals, magnetic detectors and magnetic field signals.

\subsection{Generation of multiunit multilevel droplet codes}
To prove the above concept, we choose GMR sensors as signal detectors. The GMR sensors are a type of magnetic field sensors that is capable of detecting very low magnetic fields down to a few tens of micro-Tesla \cite{Lin2017}. Ferrofluid superparamagnetic nanoparticles are used as the encoding material of droplet codes. The kind of magnetic nanoparticles, when loaded in droplets and magnetized by an external magnetic field, can generate quantifiable magnetic field patterns that can be resolved by a GMR sensor \cite{Pekas2004, Lin2014}. We chose PTFE tubes with inner diameters of a few hundred micrometers to build the fluidic networks. By connecting these tubes with pre-formed T-junctions, modular systems can be set up with ease. The millifluidic system enables to generate droplets with volumes ranging from a few picoliters to nanoliters \cite{Lin2013}. providing a wide range of droplet volumes for diverse applications to cell culturing and materials synthesis. We built a fluidic platform encompassing all the above features to flow through to enable the functionalities of code generation, code verification and sample indexing (details of the system is shown in Figure S2). 

A code is embodied by a unique sequence of droplet units and each unit should produce well-differentiable signals. Binary codes based on two discrete signal levels have been ubiquitous in nowadays information society in computing and telecommunications. Realization for binary codes is rather straightforward to follow \cite{Lin2016}, that is, by using two T-junctions to produce droplets encapsulating distinct amounts of magnetic nanoparticles (Figure S3). The droplets produced from upstream can be merged into a single channel to form an initial sequence which afterwards passes across a GMR sensor for detection. The spacing between droplets is controlled by adding spacer oils to avoid potential overlap in the signal patterns of neighboring droplets so that they will not affect the resolution. The volume of droplets may influence the signal levels when the size of droplets is smaller than the size of magnetic sensors \cite{Lin2013}. Under such a circumstance, the sensor is sensitive to both the volume of droplets and the concentrations of encapsulated magnetic particles. Hence, in current work, different concentrations of magnetic nanoparticles are explored to create distinct signal levels. We operate with droplets produced with volumes sufficient to cover the whole sensing area of the sensor so that the signal level is insensitive to the volume of droplets. Code units can be selected from the initial droplets sequence, where each droplet provides either high (pink-shaded region) or low level (green-shaded region) of magnetic signals to represent logic \lq{1}\rq{} and \lq{0}\rq{}, respectively. The combination of a certain number of \lq{1}\rq{} and \lq{0}\rq{} gives binary codes and different combinations represent different decimal numbers. For example, in 4-unit binary codes, \lq{0000}\rq{}, \lq{0001}\rq{}, \lq{0010}\rq{}, \lq{0011}\rq{},..., \lq{1111}\rq{} represent decimal numbers 0, 1, 2, 3,..., 15, respectively. Naturally, 4-unit binary codes offer an indexing capacity of 16, which is, however, not sufficient to cater for the indexing requirement in large compound libraries.

\begin{figure}
	\includegraphics[width=\linewidth]{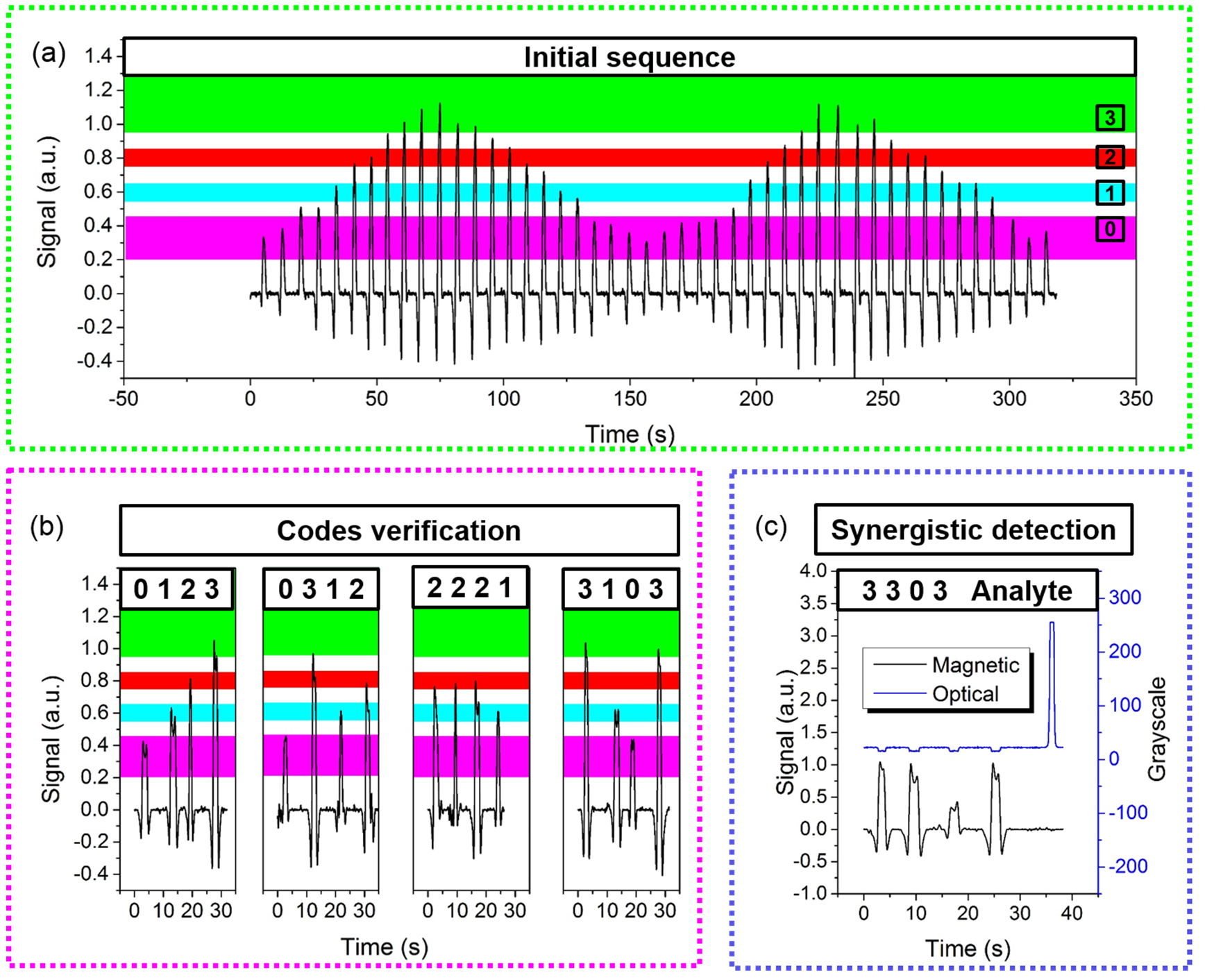}
	\caption{(a) Real-time detection signal of the initial droplet sequence for generating quaternary codes. Shaded regions represent distinct levels of a quaternary code. (b) Magnetic signals of the generated quaternary codes. Signals located in pink, blue, red and green shaded regions are defined as code units \lq{0}\rq{}, \lq{1}\rq{}, \lq{2}\rq{} and \lq{3}\rq{}, respectively. (c) The synergistic decoding of quaternary code \lq{3303}\rq{} and the optical detection of the encoded reaction droplet. The colour of the frames corresponds to the one in Figure~\ref{fig:figure1}e.}
    \label{fig:figure2}
\end{figure}

One way to expand the indexing capacity is to increase the distinguishable magnetic signal levels. For example, 4-unit quaternary codes offer an indexing capacity of 256, which is 16 times of that of the 4-unit binary codes and comparable with that offered by state-of-the-art multicolor optical encoding strategy. For this, we periodically adjust the flow rates of magnetic nanoparticles and water loaded in a droplet by using only one T-junction (Figure S4). With this method, a concentration gradient of magnetic nanoparticles can be loaded in droplets and hence a gradient of magnetic encoding signals can be generated (Figure~\ref{fig:figure2}a). We identify four regions from the detected signal patterns indicated by four color bands to represent the four distinct signal levels needed to form quaternary codes. The gap between the neighboring regions of the signal levels is determined based on the confidence intervals of the distribution of the signal level of a given concentration of magnetic nanoparticles. The width of a color band defines the range of the distribution of a group of droplets with various signal levels, where we could confidently ascribe a droplet of a particular signal level to a specific bit. The dominant source of variability in the signal level comes from the variations of the amounts of magnetic nanoparticles encapsulated in the droplets produced in a single batch. We note that the width of a color band can be defined freely as long as the gap between each color band is sufficiently larger than the confidential intervals of the distribution of signal levels of droplets loading a given concentration of magnetic nanoparticles. After selection, we show in Figure~\ref{fig:figure2}b that distinct quaternary codes can be generated with four examples. The logic of the selection module is adjusted in a way to realize distinct codes, e.g. \lq{0123}\rq{}, \lq{0312}\rq{}, \lq{2221}\rq{} and \lq{3103}\rq{}, which represent decimal numbers 27, 54, 169 and 211, respectively. To demonstrate that the generated codes can be used to index a droplet reactor, a droplet loading fluorescent dyes is inserted after the code. The entire unit (code and reaction droplet) is detected using the synergistically working optical and magnetic detectors. In this specific example the droplet reactor presented with a fluorescence signal (Figure~\ref{fig:figure2}c) is indexed with a magnetic code \lq{3303}\rq{} corresponding to a decimal number 243. The detection result is manifested in the form of several units of magnetic signal peaks that shows a certain pattern representing a decimal number followed by an optical signal peak produced by the fluorescence. Such a signal sequence reveals that both the codes and the presence of reaction product in the droplet reactor.

\subsection{Exploring a large library of codes}
\begin{figure}
	\includegraphics[width=\linewidth]{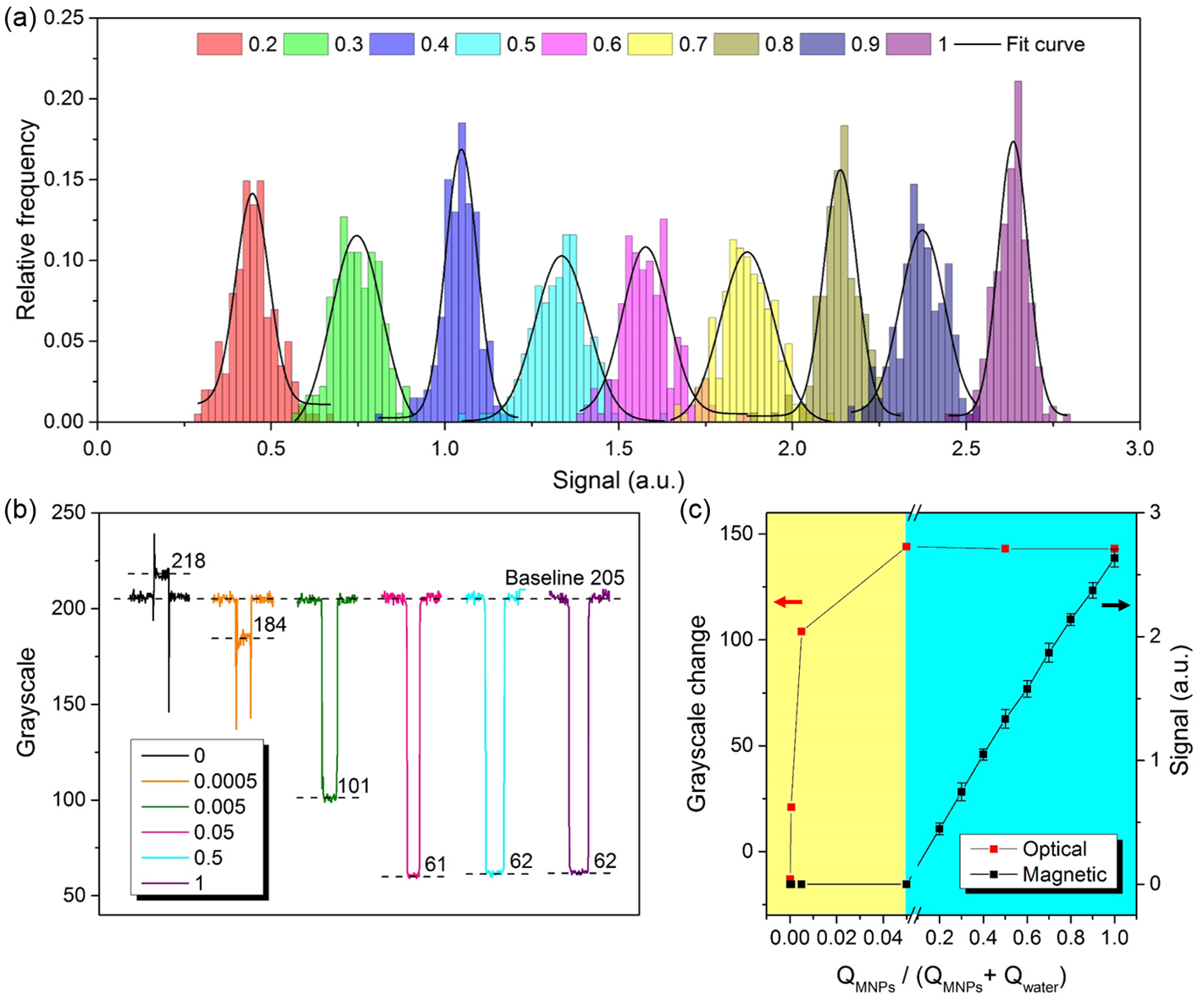}
	\caption{(a) Histograms of the distribution of magnetic signal levels for various concentrations of MNPs loaded in droplets. Around 200 data points are collected for each concentration. (b) Optical signals obtained from droplets of different MNPs concentrations. (c) Plot of optical (red points) and magnetic (black points) detection signal vs. flow parameters (proportional to MNPs concentrations) in droplets. Blue and yellow shaded regions indicate the ranges where magnetic and optical detection present discrimination capability on concentrations, respectively. Error bars indicate data ranges within one standard deviation. $Q_{\text{MNPs}} + Q_{\text{water}} = 50$ nL/s, $Q_{\text{oil}} = 200$ nL/s. Electronic sensitivity is 5 $\mu$V.}
    \label{fig:figure3}
\end{figure}

To explore the limit of indexing capacity offered using this setup and materials, a statistical analysis is performed on the magnetic signals from droplets of 9 different concentrations (Figure~\ref{fig:figure3}a), which can be clearly distinguished, suggesting an indexing capacity of 6561 can be achieved when applying 4-unit 9-level codes.
An increased indexing capacity can be further achieved by accommodating more distinguishable signal levels in the full concentration range. The average of standard deviations ($\sigma$) derived from the 9 Gaussian fits to the histograms is 0.06 a.u.. Assuming the best resolution of the GMR sensor can achieve is 3$\sigma$ ($\approx$ 0.18 a.u.), the ultimate number of distinguishable magnetic signal levels ($N$) can be estimated from the equation: 
\begin{equation}
  N = 1 + \frac{\text{signal~range}}{3 \sigma} \label{eqn:1}
\end{equation}
The signal ranges from about 0.45 a.u. to 2.64 a.u. Therefore, $N = 1 + (2.64-0.45) / 0.18 \approx 12$. With each code unit providing 12 distinguishable signal levels, 4-unit codes offer an indexing capacity of 20736. 

The number of codes is determined by the detection limit of the magnetic sensor. Still, the capacity could be extended even without changing the setup by using simple optical absorption. Droplets containing small amounts of MNPs show no magnetic signals but identifiable optical signals, i.e. when $0 < Q_{\text{MNPs}} / (Q_{\text{MNPs}} + Q_{\text{water}}) < 0.05$, exemplarily, two distinguishable optical signal levels are present (Figure~\ref{fig:figure3}b,c). In this way, 14 distinguishable states could be achieved in one droplet and the indexing capacity from 4-unit codes can be increased to 38416, which is sufficient to be used for synthesizing large compounds libraries for drug discovery applications. 

\subsection{Detection speed}

The time scale of indexing with the current system is limited by the frequency of elementary droplet bits produced from upstream and the switching rate of electromagnetic valves (about $25$ ms) used to isolate the droplets. The time scale could be improved by producing the elementary droplets initially at a high rate ($>100$ Hz) \cite{Lin2015}, which potentially allows to reduce the time scale for generating a 4-unit droplet code below $1$ s.

The velocity of droplets does not influence the signal level as long as the shape of droplets does not change, but it affects the signal width. However, droplets may deform at a high flow velocity exceeding a certain threshold due to a high shear stress. Under such a circumstance, the signal level can drop due to the shrinking of the dimensions of droplets perpendicular to the travel direction of droplets, leading to less effective magnetic stray fields collected by the sensor. In our previous study \cite{Lin2015a}, we found that when a droplet shrinks in the lateral direction by a strong magnetic force, the sensor area exposed to the droplet decreases, which is the same as the situation when they are stretched by high shear stress at a high flow velocity.  The voltage signal follows a scaling rule for a constant ferrofluid concentration: $V \propto S$, where $S$ is the sensor area exposed to ferrofluid droplets. Hence, with a constant fluctuation of concentration $\Delta C$, the fluctuation of voltage signal $\Delta V$ decreases with the decrease of $S$. Furthermore, droplets encapsulating the highest concentration of ferrofluid (Ferrotec EMG 700, commercial stock concentration: $75$ mg/mL) determines the whole signal range ($V_{\text{S}}$), which decreases with the exposed sensor area $S$. As the number of resolvable different signal levels ($N$) is determined by $V_{\text{S}}/\Delta V$. In this respect, two scenarios can apply: (i) When $\Delta V$ is larger than the limit of detection of the sensor, the number of distinct signal levels that can be generated is not changed because both $V_{\text{S}}$ and $\Delta V$ decrease proportionally. (ii) When $\Delta V$ is smaller than the limit of detection of the sensor, the noise of detection dominates over the signal change caused by the fluctuation of the concentration of ferrofluid. In this case, the number of distinct signal levels will decrease accordingly. However, in this scenario, solidifying droplets into solid particles would mitigate this effect. This requires the production of droplets with volumes smaller than the channel cross section to avoid channel clogging and using ferrofluid blended with photosensitive polymers that can be crosslinked by UV illumination \cite{Zhao2012}.

\section{Conclusions}

Droplet microfluidic devices are promising platforms for large-scale biological assays and combinatorial synthesis. The capability to identify each droplet reactor is crucial towards the realization of high-throughput biological assays, custom multistep synthesis in a fully automated way. We have reported a multiunit multilevel positional encoding strategy to encode droplet reactors, with each droplet unit possessing several distinct signal levels. For the proof-of-concept demonstration, a millifluidics platform armed with an encoding functionality and a synergistic optical and magnetic detection system was built up and a sequence of a reaction droplet encoded using a 4-unit quaternary code was created and decoded using the devices.

The capacity limit using the setup and materials was investigated. Using only magnetic signals, 12 distinguishable magnetic levels in one code unit can be achieved, therefore the indexing capacity of 4-unit codes can reach 20736. By further adding optical signal as an encoding dimension, the indexing capacity of codes can reach around 38416. The proposed encoding strategy is advantageous in terms of its potential to encoding a large number of encoding capacity without scaling up the system compared with other current encoding methods such as multi-colour encoding. The positional encoding method is also beneficial for droplet-based screening that excludes the possibility of sample contamination due to the presence of encoding materials in the same reactors. We anticipate that if the sensitivity and resolution of the magnetic sensor can be further improved or the concentration and the magnetic moments of magnetic nanoparticles can be increased, more distinguishable signal levels will be available. Encoding capacity exceeding $10^6$, which is of the scale of the size of some large compound libraries, is expected to be achieved.

\section{Experimental}

\subsection{The measurement platform}
The integrated setup is composed of 4 functional modules (Figure S2): (a) droplet generating module, (b) optical detection module, (c) magnetic detection module and (d) sorting module.

\subsection{Creation of droplets}
The droplet generator module (Figure S2a) was assembled from: the pulse free syringe pumps (neMESYS, Cetoni GmbH) for fluid injection; syringes (1000 Series, Hamilton, syringe volume: 1.0 mL and 2.5 mL); the T-junctions (IDEX, ETFE 1/16" tubing, thru-hole diameter: 0.5 mm) for droplet formation; flangeless fittings (standard 1/4"--28 thread) for interconnection between junctions and tubing, flangeless ferrule (IDEX, ETFE, Blue, 1/16" OD tubing, 1/4"--28 Flat Bottom); polytetrafluoroethylene (PTFE) tubing with outer diameter of 0.9 mm and inner diameter of 0.5 mm that realize the flow circuits.

To gain magnetic functionality, we rely on commercial available ferrofluid (water-based magnetic nanoparticles solution, Ferrotec EMG Series 700, concentration: 5.8\% vol.). The MNPs droplets were formed in T-junctions, where MNPs solution was dispersed in the oil phase (FC-40™ Fluorinert™, 3M). Each created droplet was characterized by length and width as indicated in Figure S2, Inset 1. Droplets are produced that the sizes thereof are larger than the sensor size (completely cover the whole sensing area) to avoid the influence of droplet size on the sensor signal. The spacing between droplets is adjusted by controlling the flow rate of spacer oils to physically separate neighboring droplets. For current work, the flow velocity is determined by the total flow rates of oil and ferrofluid injected from upstream to produce stable droplet trains, as the whole system is operated in a continuous flow. We chose flow rates so that droplets were produced stably and completely filling the channel cross-section. The flow velocity could be eventually adjusted by changing the upstream flow rates. The concentration of ferrofluid is chosen so that they fall in the linear sensing range of the magnetic sensor, as shown in Figure~\ref{fig:figure3}c.

The magnetization of ferrofluid droplets could be affected by other droplets when the droplets are getting sufficiently close to each other because of the influence of magnetic dipole field. This can be reflected by overlapped magnetic detection signal patterns, which were not seen in our experiments (Figure~\ref{fig:figure2}). In current experiments, we used an extra channel to inject spacer oil in between droplets so that the stray field of neighboring droplets do not interfere each other, which is indicated by the isolated detection peaks. In current work, we don’t observe that the magnetization of droplets is affected by the flow speed.

To create fluorescein droplets, 0.1 g 5(6)-carboxyfluorescein (Sigma-Aldrich GmbH) was dissolved in 20 mL 0.1 mol/L phosphate-buffered saline (PBS) solution as disperse phase. Perfluorinated silicone oil (FC-40™ Fluorinert™, 3M) was used as the continuous phase. 

\subsection{Optical detection module} 
A microscope (Zeiss, Scope. A1) equipped with a high-speed camera (Photron, Fastcam SA3, black and white imaging) was used as the optical detection module (Figure S2b). An objective lens with 5x magnification was applied. The microscope was operated in bright-field mode while detecting magnetic droplets. When the fluorescent light needs to be collected, the microscope was operated with a fluorescence filter. 
The analog output from the camera was transferred to the computer. A LabView program was used to extract the center pixel grayscale of the live camera image. The grayscale ranges from 0 to 255, with 0 as pure dark and 255 as pure bright. When a dark object passes the detection spot, the grayscale of the image decreases; when a luminescent object passes the detection spot, the grayscale increases. Therefore, when the dark ferrofluid droplets passed the probe light spot, the grayscale of the center pixel decreased. As a consequence, rectangular-shape peaks (Figure S2b, Inset 2) formed when a train of droplets passes by. 

\subsection{Magnetic detection module} 
The magnetic detection was realized with a GMR sensor. The layer stack of the GMR multilayers was: [Py(1.5 nm) / Cu(2.2 nm)]$_{30}$ / Py(1.5 nm), (Py = Ni$_{81}$Fe$_{19}$) \cite{Lin2016}.  GMR sensors are proximity sensors, which response to local magnetic stray fields from the objects. Therefore, a strategy is taken to enhance the magnetic stray fields to be detected. As the ferrofluid employed for present study is a suspension of magnetic colloidal nanoparticles in water, it does not retain magnetization without applying external magnetic field. Thus, an external field should be applied to achieve a net magnetization of the ferrofluid. In addition, the external magnetic field can be simultaneously used to bias the sensor to the sensitive region. Based on these considerations, a permanent magnet (AlNiCo 500, type A1560, IBSMagnet, length: 60 mm, diameter: 15 mm) was placed below the GMR sensor. The position of the permanent magnet was adjusted to achieve detectable stray field from MNPs droplets and to bias the GMR sensor to the most sensitive region. 

The GMR sensor was connected with 3 trimmers (Bourns, 3006P-1-102LF) to realize a Wheatstone bridge geometry (Figure S2c). The measured differential output voltage from the bridge scales with the change of external magnetic field \cite{Lin2016}. By balancing the bridge, the background voltage across the bridge can be minimized so that the measuring sensitivity can be enhanced. 

A lock-in amplifier (AMETEK®, Model 7230) was used to amplify the voltage across the bridge and reduce the noise. The whole bridge circuit was also powered by the lock-in using a constant AC voltage of 0.2 V. The modulation frequency and the time constant were set as 1 kHz and 100 ms, respectively. The line filter is set to filter 50 Hz and 100 Hz noise. The analog output from the lock-in was converted to digital data through a data acquisition box (NI, USB-6009). A LabView program was used to visualize and analyze the digital data. The characteristic peaks (Figure S2c, Inset 3) formed when a train of droplets pass by. The peak amplitude and width were determined instantly by the LabView program. This information was used as a feedback for automatically triggering the switch of a 3-way valve, relying on which droplets of interest were sorted into the codes container (Figure S2d).

\subsection{Sorting module}
The sorting function was realized by switching the 3-way valve (Figure S2d). A solenoid 3-way isolation valve (Amico Scientific) was with one way open and the other way closed without supplying power. Once a voltage of 20 V and a compliance current of 0.5 mA (provided by Agilent Precision Source/Measurement Units B2902A) was applied to the solenoid valve, the open and the closed ways were switched. 

Magnetically activated sorting was used for generating magnetic codes. The process of magnetic codes generation is as follows: Voltage ranges were defined for different code bits. MNPs droplets of different concentrations are detected by the GMR sensor. The detected signal was analyzed and compared with the pre-defined voltage ranges. Once droplets of interest were detected, the solenoid 3-way isolation valve was switched to allow them to enter the codes container. A release-time in the LabView program is set to determine how much time a droplet needs to travel from the GMR sensor to the sorting Y-junction. In this way, at which time and how long the valve should be kept open is calculated by the LabView program. This feature is crucial for the precise control of the trajectory of droplets. The injection of fluorescein fluid is carried out by manually triggering the pumping program in the computer.

\subsection{Generation of the initial sequence for quaternary codes}
Figure S4a and Table~\ref{tbl:table1} show the schematic of the fluidic circuit and the flow rate program that were used to generate the initial sequence for quaternary codes. The generated train of droplets presented a regular signal pattern, with 160 s as a period and in each period the signal level first increase and then decrease (Figure S4b). The sensitivity of the lock-in was set to 10 $\mu$V.

\begin{table}
  \caption{One cycle of the flow rate program for generating the initial sequence for the quaternary codes. $Q_{\text{oil}} = 200$ nL/s.}
  \label{tbl:table1}
  \begin{tabular}{ccc}
    \hline
    Time (s) & $Q_{\text{MNPs}}$ (nL/s) & $Q_{\text{water}}$ (nL/s) \\
    \hline
    0--20   & 0.0 & 50.0 \\
    20--40 & 12.5 & 37.5\\
    40--60  & 25.0 & 25.0\\
    60--80 & 37.5 & 12.5\\
    80--100 & 50.0 & 0.0\\
    100--120 & 37.5 & 12.5\\
    120--140 & 25.0  & 25.0\\
    140--160 & 12.5 & 37.5\\
     \hline
  \end{tabular}
\end{table}

\begin{acknowledgement}

We acknowledge insightful discussions with Dr. Larysa Baraban (TU Dresden). We thank T. Voitsekhivska, G. S. Canon Bermudez, B. Scheumann (all from HZDR) for the assistance in sample fabrication and characterization. Support by the Nanofabrication Facilities Rossendorf and Structural Characterization Facilities Rossendorf at the Ion Beam Center (IBC) at the HZDR is greatly appreciated. The work was financially supported in part via the German Science Foundation (DFG) Grant MA 5144/9-1 and the European Research Council under the European Union’s Seventh Framework Programme (FP7/2007-2013) / ERC Grant Agreement No. 306277 and ERC Proof-of-Concept grant funded via the European Union's Horizon 2020 research and
innovation programme (Grant Agreement No. 768584).

\end{acknowledgement}

\begin{suppinfo}

The Supporting Information file contains further details on the experimental setup, the formation of binary codes and the envisioned application example. The Supporting Information is provided as a single PDF file.

\end{suppinfo}


%

\newpage
\section{for TOC only}
\begin{figure}
	\includegraphics[height=100px]{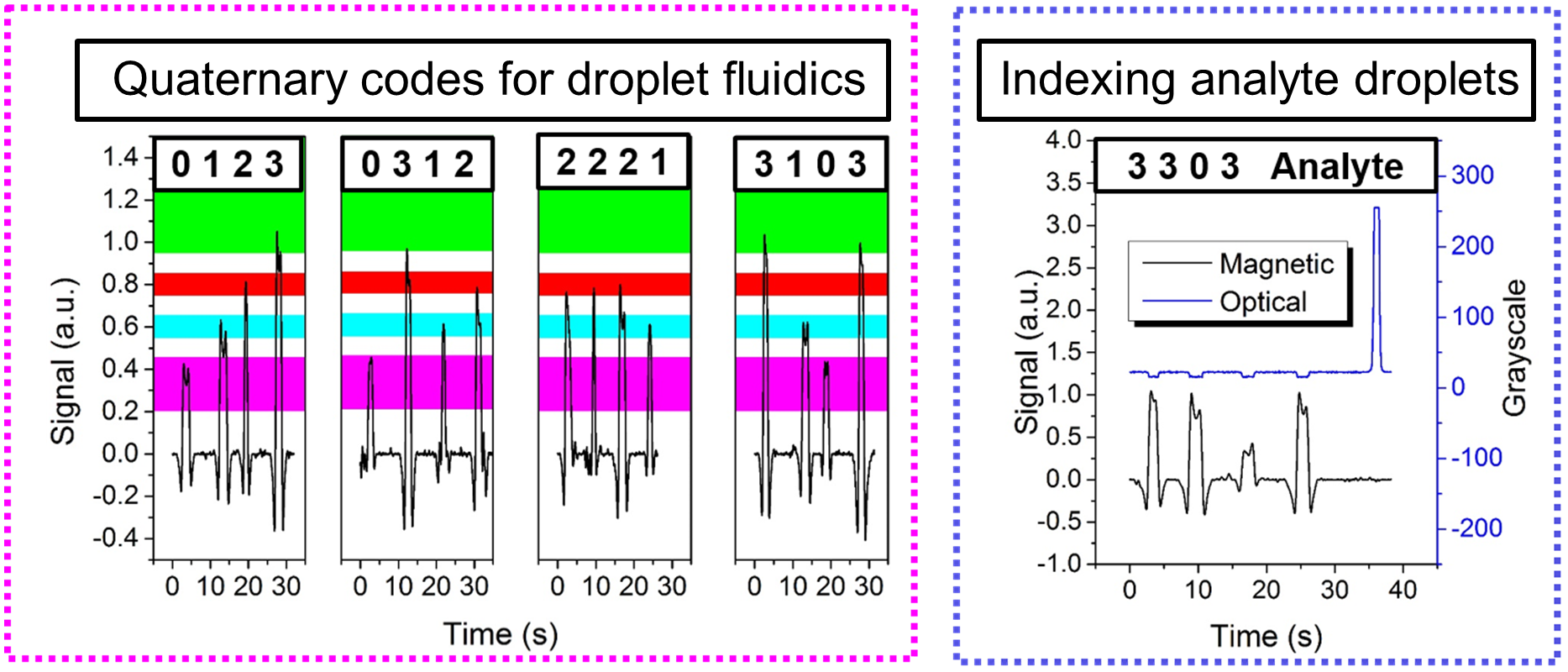}
\end{figure}

\end{document}